\documentclass{article}
\usepackage{amssymb}
\usepackage{amsfonts}
\usepackage{amsmath}

\begin{document}

\title{Group-theoretical approach to a non-central extension of the Kepler-Coulomb
problem}
\author{G. A. Kerimov$^{1}$ and A. Ventura$^{2,3}$ \\
$^{1}$\textit{Physics Department, Trakya University, 22030 Edirne, Turkey}\\
$^{2}$\textit{ENEA, Centro Ricerche Ezio Clementel, Bologna, Italy}\\
$^{3}$\textit{Istituto Nazionale di Fisica Nucleare, Sezione di Bologna,
Italy}}
\maketitle

\begin{abstract}
Bound and scattering states of a non-central extension of the
three-dimensional Kepler-Coulomb Hamiltonian are worked out analytically
within the framework of the potential groups of the problem, $SO\left(
7\right) $ for bound states and $SO(6,1)$ for scattering states. In the
latter case, the $S$ matrix is calculated by the method of intertwining
operators.
\end{abstract}

\section{Introduction}

In classical mechanics, the reduced Kepler problem has been known for more
than two centuries \cite{Go75} to admit seven integrals of motion.\textit{\ }%
These are the total angular momentum, the Laplace-Runge-Lenz (or
Hermann-Bernoulli-Laplace ) vector and the total energy. Since there are two
relationships between them ( see, for example, \cite{GPS01} ) only five of
integrals of motion are independent. In general, a closed system with $N$
degrees of freedom can have at most $2N-1$ independent integrals of motion 
\cite{LL76}. According to the Liouville theorem, the system is completely
integrable if it allows $N$\ integrals of motion (including the Hamiltonian
) that are independent and in involution ( i.e. Poisson brackets of any two
integrals are zero). The system is called superintegrable if there exist $q$%
, $1\leq q\leq N-1$, additional independent integrals of motion. The cases $%
q=1$\ and $q=N-1$\ correspond to minimal and maximal superintegrability,
respectively. In quantum mechanics the definitions of complete integrability
and superintegrability are same, but Poisson brackets are replaced by
commutators.

The first systematic search for quantum integrable one-particle systems with
scalar potentials was begun by Smorodinsky and co-workers in \cite{Fr65,
Wi67, Ma67} and continued by Evans in \cite{Ev90}. It was restricted to the
cases when integrals of motion are first- or second -order polynomials in
the momenta. They found all superintegrable systems in two and three
dimensions with at most second order integrals of motion. It turns out that
they possess properties making them of special interest: for instance, all
these potentials admit the separation of variables in several coordinate
systems and possess dynamical symmetries responsible for the separability of
the Schr\"{o}dinger equation. The history of this problem and some results
may be found in \cite{Wi09}.

It is well known that the first quantum study of the hydrogen atom \cite
{Pa26} was based upon the algebra generated by integrals of motion, before
the Schr\H{o}dinger equation was published. Later on, Fock \cite{Fo35} and
Bargmann \cite{Ba36} recognized that the angular momentum and the
Laplace-Runge-Lenz vector generate the Lie algebra of $SO(4)$\ in the
subspace of \ negative energies and the Lie algebra of $SO(3,1)$\ in the
subspace of \ positive energies. It was realized that the `accidental'
degeneracies, i.e. degeneracies not connected with geometrical $SO(3)$
symmetries of the Hamiltonian, are due to the invariance group $SO(4)$.
Moreover, the separation of variables in parabolic coordinates was related
to Laplace-Runge-Lenz vector\cite{Ba36}. Later on, Zwanziger \cite{Zw67}
showed that the algebra of $SO(3,1)$ may be used to calculate the Coulomb
phase-shifts. Ever since, invariance algebras have been determined for many
quantum mechanical systems. The best known of these systems are the
oscillator \cite{JH40} and the MICZ-Kepler system\cite{Zw68, MC70}. This is
a situation in which the Hamiltonian $H$ of the system belongs to the centre
of the enveloping algebra of some group $G$, i.e. 
\begin{equation}
H=f(C),  \label{H_Ci}
\end{equation}
where $C$ is the Casimir operator of the invariance group $G$. For example,
in the Coulomb bound-state problem, $H=-\gamma ^{2}/2(C+1)$, where $C$ is a
Casimir operator of $SO(4)$.

But it could happen that the Hamiltonian $H_{\nu }$\ can be related to the
Casimir operator $C$ as 
\begin{equation}
H_{\nu }=\left. f\left( C\right) \right| _{\mathfrak{H}_{\nu }}\;,
\label{H_Cp}
\end{equation}
where $\mathfrak{H}_{\nu }$\ a subspace occurring in the subgroup reduction
and $\left. {}\right| _{\mathfrak{H}_{\nu }}$ denotes the restriction to $%
\mathfrak{H}_{\nu }$. In this case the group $G$\ describes the same energy
states of a family of Hamiltonians $H_{\nu }$ with different potential
strength. (This is why the present group $G$\ designated potential group 
\cite{AGI83}.) Such an approach was proposed by Ghirardi \cite{Gh72}, who\
worked it out in detail for the Scarf potential\textbf{\ }\cite{Sc58}. It is
similar to the approach of Olshanetsky and Perelomov \cite{OP77, OP83},
where quantum integrable systems are related to radial part of the Laplace
operator on homogeneous spaces \textbf{(}i.e. to radial part of Casimir
operator of second order\textbf{)} of Lie groups.

Ref. \cite{Ke98} proposed a method that permits purely algebraic
calculations of $S$-matrices for the systems whose Hamiltonians are related
to the Casimir operators $C$ of some Lie group $G$ as (\ref{H_Ci}) or (\ref
{H_Cp}). Namely, the $S$-matrices for the systems under consideration are
associated with intertwining operators $A$ between Weyl equivalent
representations $U^{\chi }\,$and$\,\ U^{\overset{\thicksim }{\chi }}$ of $G$
as 
\begin{equation}
S=A  \label{assoc_1}
\end{equation}
or 
\begin{equation}
\text{\ }S=\left. A\right| _{\mathfrak{H}_{\nu }}  \label{assoc_2}
\end{equation}
respectively. ( The representations $U$ and $U^{\overset{\thicksim }{\chi }}$
have the same Casimir eigenvalues. Such representations are called Weyl
equivalent.) At this stage we note that the operator $A$ is said to
intertwine the representations $U^{\chi }\,$and$\,\ U^{\overset{\thicksim }{%
\chi }}$ of the group $G$\ if the relation 
\begin{equation}
AU^{\chi }(g)=U^{\overset{\thicksim }{\chi }}(g)A\text{ \ \ for\thinspace
all }g\in G  \label{int_1}
\end{equation}
or 
\begin{equation}
AdU^{\chi }(b)=dU^{\overset{\thicksim }{\chi }}(b)A\text{ \ \ for\thinspace
all}\ b\in \mathfrak{g}  \label{int_2}
\end{equation}
holds, where $dU^{\chi }$ and $dU^{\widetilde{\chi }}$ are the corresponding
representations of the algebra $\mathfrak{g}$ of $G$. Equations (\ref{int_1}%
) and (\ref{int_2}) have much restriction power, determining the
intertwining operator up to a constant.

The potential group approach has been proven to be useful in variety
problems in one dimension. Recently, it has been used to describe some
potentials \cite{KE06, Ke06, Ke07, KE07} classified in \cite{Ma67}\textbf{. }%
In Ref. \cite{Ke06} it has been shown that the superposition of the Coulomb
potential with one barrier term \cite{Ma67} could be related to the
potential group $SO(5)$. Scattering amplitudes for such system are worked
out in detail in Ref.\cite{Ke07} by using an intertwining operator \cite
{Ke98} between two Weyl-equivalent unitary irreducible representations of
the $SO(5,1)$ potential group.

Subject of the present work will be the simultaneous description of bound
and scattering states of a quantum mechanical system with Hamiltonian 
\begin{equation}
H=-\frac{1}{2}\nabla ^{2}-\frac{\gamma }{r}+\frac{s_{1}^{2}-1/4}{2x^{2}}+%
\frac{s_{2}^{2}-1/4}{2y^{2}}+\frac{s_{3}^{2}-1/4}{2z^{2}}  \label{H_3_nc}
\end{equation}
written in units $\hbar =m=1$, where $s_{i}=0,1,2,\ldots $.\ We show that

\begin{equation}
H=-\frac{\gamma ^{2}}{\left. 2(C+\frac{25}{4})\right| _{\mathcal{H}%
_{s_{1}s_{2}s_{3}}}}  \label{H_C}
\end{equation}
where $C$ \ is a Casimir operator of $SO(7)$ (for bound states) or $SO(6,1)$
(for scattering states).

This system was proved to be minimally superintegrable \cite{Ma67}, since
four integrals of motions were explicitly derived, as a consequence of the
separability of the related Schr\H{o}dinger equation in two coordinate
systems. But in Ref. \cite{VE08} it has been shown that the classical
counterpart of Hamiltonian (\ref{H_3_nc}) is maximally superintegrable, i.e.
it admits five independent integrals of\ motion, including the Hamiltonian:
four of them derive from separability of the related Hamilton-Jacobi
equation in different coordinate systems, but the fifth integral, first
discussed in Ref.\cite{VE08}, is not connected with separability. Moreover,
this last integral is quartic in the momenta, while the other three are
quadratic, and has been rederived in Ref.\cite{Ro09} as an example of
application of a more general technique.

\section{General formalism}

Let us start the discussion with the fact that the generators of UIR of $%
SO\left( 7\right) $ ( or $SO\left( 6,1\right) $) are 21 independent
Hermitian operators $M_{\mu \nu }=-M_{\nu \mu }$ ($\mu ,\nu =1,2,\ldots ,7$)
which obey the commutation relations 
\begin{equation}
\left[ M_{\mu \nu },M_{\sigma \lambda }\right] ={\normalsize i}\left( g_{\mu
\sigma }M_{\nu \lambda }+g_{\nu \lambda }M_{\mu \sigma }-g_{\mu \lambda
}M_{\nu \sigma }-g_{\nu \sigma }M_{\mu \lambda }\right)  \label{comm_rel}
\end{equation}
where 
\begin{eqnarray}
g_{\mu \nu } &=&\left( +,+,\ldots ,+,+\right) \quad \text{for \ \ }SO\left(
7\right)  \label{metric} \\
g_{\mu \nu } &=&\left( +,+,\ldots ,+,-\right) \quad \text{for \ \ }SO\left(
6,1\right)  \notag
\end{eqnarray}

There are three independent Casimir invariants which are identically
multiple of the unit in each UIR. In the case of most degenerate
representations, they are identically zero, with the exception of the second
order Casimir operator 
\begin{equation}
C=\frac{1}{2}\sum\limits_{\mu ,\nu =1}^{7}M_{\mu }^{\nu }M_{\nu }^{\mu }\;,
\label{C_a}
\end{equation}

It is well-known that the most degenerate representation of algebra $so(7)$
( $so(6,1)$ ) can be realized in the Hilbert space spanned by
negative-energy (positive-energy) states corresponding to fixed eigenvalue
of the Coulomb Hamiltonian $H^{Coul}$\ in six dimensions 
\begin{equation}
H^{Coul}=\frac{1}{2}p^{2}-\frac{\gamma }{\sqrt{x^{2}}},\text{ \ \ }\gamma >0
\label{Ham_C}
\end{equation}
where $x=\left( x_{1},x_{2},\ldots ,x_{6}\right) \in R^{6}$, $\ p_{j}=-i%
\frac{\partial }{\partial x_{j}}$, ($j=1,...,6$), $x^{2}=%
\sum_{i=1}^{6}x_{i}x_{i}$, $p^{2}=\sum_{i=1}^{6}p_{i}p_{i}$. (We are using
units with $M=\hbar =1$.)\ However, in order to be able to write the
relation (\ref{H_Cp}) we introduce the following realization 
\begin{equation}
M_{ij}=\lambda \left( x\right) \circ \left( x_{i}p_{j}-x_{j}p_{i}\right)
\circ \lambda ^{-1}\left( x\right)  \label{gen_1}
\end{equation}
\begin{equation}
M_{i7}=-M_{7i}=\left\vert 2h\right\vert ^{-\frac{1}{2}}\lambda \left(
x\right) \circ \left[ x_{i}p^{2}-p_{i}\left( x\cdot p\right) +{\normalsize i}%
\frac{5}{2}p_{i}-\frac{\gamma x_{i}}{x^{2}}\right] \circ \lambda ^{-1}\left(
x\right) ,\;(\ i,j=1,...,6)  \label{gen_2}
\end{equation}
where

\begin{equation}
\lambda \left( x\right) =\left[ \left( x_{1}^{2}+x_{2}^{2}\right) \left(
x_{3}^{2}+x_{4}^{2}\right) \left( x_{5}^{2}+x_{6}^{2}\right) \right] ^{1/4}\;
\label{lambda_r}
\end{equation}
and 
\begin{equation}
h=\lambda \left( x\right) \circ \left( \frac{1}{2}p^{2}-\frac{\gamma }{\sqrt{%
x^{2}}}\right) \circ \lambda ^{-1}\left( x\right) .  \label{h}
\end{equation}
The generators (\ref{gen_1}-\ref{gen_2}\textbf{) }act in the eigenspace of $%
h $\ equipped with the scalar product 
\begin{equation}
\left( \phi _{1},\phi _{2}\right) =\int\limits_{R^{6}}\phi _{1}^{\ast
}\left( x\right) \phi _{2}\left( x\right) d\mu \left( x\right) ,\quad x\in
R^{6}  \label{scal_prod}
\end{equation}
where $d\mu \left( x\right) =\lambda ^{-2}\left( x\right) dx_{1}dx_{2}\cdots
dx_{6}.$

This representation, of course, is unitarily equivalent to the
representation constructed in the eigenspace of the Coulomb Hamiltonian $%
H^{Coul}$\ in six dimensions.\ The unitary mapping $W$ which realizes the
equivalence is given by 
\begin{equation}
W:\text{ \ \ \ }\Psi ^{Coul}\rightarrow \Phi =\lambda \left( x\right) \Psi
^{Coul}  \label{U_E}
\end{equation}

The operators (\ref{gen_1}-\ref{gen_2}) provide most degenerate
representations of $SO\left( 7\right) $ if $h$\ is negative definite and of $%
SO\left( 6,1\right) $ if $h$\ is positive definite. More precisely, they
define the most degenerate (symmetric) UIR of $SO\left( 7\right) $ specified
by the integer number $j=0,1,\ldots $(when $h$\ is negative definite) and
the most degenerate principal series representations of $SO\left( 6,1\right) 
$ labelled by the complex number $j=-\frac{5}{2}+i\rho ,\ \rho >0$ (when $h$%
\ is positive definite). \ If we compute the second-order Casimir operator (%
\ref{C_a}), it becomes 
\begin{equation}
C=-\frac{25}{4}-\frac{\gamma ^{2}}{2h}  \label{C_b}
\end{equation}

Let us consider the reduction corresponding to the group chain $G\supset
SO\left( 6\right) \supset SO\left( 4\right) \times SO\left( 2\right) \supset
SO\left( 2\right) \times SO\left( 2\right) \times SO\left( 2\right) ,$\
where $G$\ is $SO\left( 6,1\right) $\ or $SO\left( 7\right) $. Then, the
basis functions can be characterized by the Casimir operators of \ the chain
of groups 
\begin{eqnarray}
C\left| j;lM\right\rangle &=&j\left( j+5\right) \left| j;lM\right\rangle
\label{eigen_eqs} \\
C^{SO(6)}\left| j;lM\right\rangle &=&l\left( l+4\right) \left|
j;lM\right\rangle  \notag \\
C^{SO(4)}\left| j;lM\right\rangle &=&m\left( m+2\right) \left|
j;lM\right\rangle  \notag \\
C^{SO(2)_{1}}\left| j;lM\right\rangle &=&s_{1}^{2}\left| j;lM\right\rangle 
\notag \\
C^{SO(2)_{2}}\left| j;lM\right\rangle &=&s_{2}^{2}\left| j;lM\right\rangle 
\notag \\
C^{SO(2)_{3}}\left| j;lM\right\rangle &=&s_{3}^{2}\left| j;lM\right\rangle 
\notag
\end{eqnarray}
where $M$ is a collective index$\ \left( m,s_{1},s_{2},s_{3}\right) $ and 
\begin{equation}
C^{SO(6)}=\frac{1}{2}\sum_{i,j=1}^{6}M_{ij}^{2},\ C^{SO(4)}=\frac{1}{2}%
\sum_{i,j=1}^{4}M_{ij}^{2},C^{SO(2)_{1}}=M_{12}^{2},C^{SO(2)_{2}}=M_{34}^{2},C^{SO(2)_{3}}=M_{56}^{2}
\label{C_6_4}
\end{equation}

According to this, we introduce in place of $x_{1},x_{2},\ldots ,x_{6}$ the
variables $r,\theta ,\varphi ,\alpha _{1},\alpha _{2},\alpha _{3}$\ via $%
x_{i}=rn_{i}$ with 
\begin{equation}
\begin{array}[t]{l}
n_{1}=\sin \theta \sin \varphi \sin \alpha _{1},\quad n_{2}=\sin \theta \sin
\varphi \cos \alpha _{1} \\ 
n_{3}=\sin \theta \cos \varphi \sin \alpha _{2},\quad n_{4}=\sin \theta \cos
\varphi \cos \alpha _{2} \\ 
n_{5}=\cos \theta \sin \alpha _{3},\qquad \ \ \ \ n_{6}=\cos \theta \cos
\alpha _{3}
\end{array}
\label{polar_coord}
\end{equation}
where $0\leq r<\infty ,\quad 0\leq \theta ,\varphi \leq \frac{\pi }{2}$ and $%
0\leq \alpha _{1},\alpha _{2},\alpha _{3}\leq 2\pi $. If we compute the
operator $\gamma ^{2}/\left( C+\frac{25}{4}\right) $ for this
parametrization, it becomes 
\begin{eqnarray}
\frac{\gamma ^{2}}{C+\frac{25}{4}} &=&\frac{\partial ^{2}}{\partial r^{2}}+%
\frac{2}{r}\frac{\partial }{\partial r}+\frac{1}{r^{2}}\left( \frac{1}{\sin
\theta }\frac{\partial }{\partial \theta }\sin \theta \frac{\partial }{%
\partial \theta }+\frac{1}{\sin ^{2}\theta }\frac{\partial ^{2}}{\partial
\varphi ^{2}}\right)  \label{h'} \\
&&+\frac{1}{r^{2}\sin ^{2}\theta \sin ^{2}\varphi }\left( \frac{1}{4}+\frac{%
\partial ^{2}}{\partial \alpha _{1}^{2}}\right) +\frac{1}{r^{2}\sin
^{2}\theta \cos ^{2}\varphi }\left( \frac{1}{4}+\frac{\partial ^{2}}{%
\partial \alpha _{2}^{2}}\right) +\frac{1}{r^{2}\cos ^{2}\theta }\left( 
\frac{1}{4}+\frac{\partial ^{2}}{\partial \alpha _{3}^{2}}\right)  \notag
\end{eqnarray}
\qquad \qquad

Let $\mathcal{H}_{s_{1}s_{2}s_{3}}$ be a subspace spanned by $\left\vert
j;lM\right\rangle $ with fixed $s_{1}$, $s_{2}$ and $s_{3}$. Thus, the
operator (\ref{h'}) restricted to this subspace becomes a differential
operator in $r$, $\theta $ and $\varphi $; it turns out that 
\begin{eqnarray}
\left. \frac{\gamma ^{2}}{C+\frac{25}{4}}\right\vert _{\mathcal{H}%
_{s_{1}s_{2}s_{3}}} &=&\frac{\partial ^{2}}{\partial r^{2}}+\frac{2}{r}\frac{%
\partial }{\partial r}+\frac{1}{r^{2}}\left( \frac{1}{\sin \theta }\frac{%
\partial }{\partial \theta }\sin \theta \frac{\partial }{\partial \theta }+%
\frac{1}{\sin ^{2}\theta }\frac{\partial ^{2}}{\partial \varphi ^{2}}\right)
\\
&&+\frac{1/4-s_{1}^{2}}{r^{2}\sin ^{2}\theta \sin ^{2}\varphi }+\frac{%
1/4-s_{2}^{2}}{r^{2}\sin ^{2}\theta \cos ^{2}\varphi }+\frac{1/4-s_{3}^{2}}{%
r^{2}\cos ^{2}\theta }  \notag
\end{eqnarray}
with $s_{i}=0,\pm 1,\pm 2\ldots $, where we have used\ that 
\begin{equation*}
C^{SO(2)_{i}}=-\frac{\partial ^{2}}{\partial \alpha _{i}^{2}},i=1,2,3
\end{equation*}
Hence, the Hamiltonian 
\begin{eqnarray}
H &=&-\frac{1}{2}\nabla ^{2}-\frac{\gamma }{r}+\frac{s_{1}^{2}-1/4}{%
2r^{2}\sin ^{2}\theta \sin ^{2}\varphi }  \label{h_gen_coul} \\
&&+\frac{s_{2}^{2}-1/4}{2r^{2}\sin ^{2}\theta \cos ^{2}\varphi }+\frac{%
s_{3}^{2}-1/4}{2r^{2}\cos ^{2}\theta }  \notag
\end{eqnarray}
can be described in terms of the potential groups $SO\left( 7\right) $ and $%
SO\left( 6,1\right) $ since 
\begin{equation*}
H=-\left. \frac{\gamma ^{2}}{C+\frac{25}{4}}\right\vert _{\mathcal{H}%
_{s_{1}s_{2}s_{3}}}\;,
\end{equation*}
as mentioned in Section 1 (formula (\ref{H_C})). (Due to the symmetry $%
s_{i}\rightarrow -s_{i}$ in the Hamiltonian (\ref{h_gen_coul}), without\
loss of generality, we may assume that $s_{1}$, $s_{2}$ and $s_{3}$ are
non-negative integers.) Note that the $SO(2)$ subgroups are related to
potential strength.

At this point, it is worthwhile pointing out that Hamiltonian (\ref
{h_gen_coul}) does not contain the pure Coulomb potential as a particular
case, within the framework of the $SO\left( 7\right) $\ and $SO\left(
6,1\right) $\ symmetries considered in the present work. In order to restore
it, it is necessary to resort to larger symmetry groups, for example, $%
SO\left( 10\right) $\ and $SO\left( 9,1\right) $\ and use the decomposition
chain $G\supset SO\left( 9\right) \supset SO\left( 6\right) \times SO\left(
3\right) \supset SO\left( 3\right) \times SO\left( 3\right) \times SO\left(
3\right) ,$\ where now the $SO(3)$ subgroups are related to potential
strength.

Here again, use is made of polar coordinates 
\begin{equation*}
x=\left( \sin \theta \sin \varphi e_{1},\sin \theta \cos \varphi e_{2},\cos
\theta e_{3}\right)
\end{equation*}
where $x\in R^{9},$ $e_{i}=\left( \sin \alpha _{i}\sin \beta _{i},\sin
\alpha _{i}\cos \beta _{i},\cos \alpha _{i}\right) ,i=1,2,3$. Then, a
procedure similar to that described above would lead to the Hamiltonian 
\begin{equation*}
H=-\frac{1}{2}\nabla ^{2}-\frac{\gamma }{r}+\frac{l_{1}(l_{1}+1)}{2r^{2}\sin
^{2}\theta \sin ^{2}\varphi }+\frac{l_{2}(l_{2}+1)}{2r^{2}\sin ^{2}\theta
\cos ^{2}\varphi }+\frac{l_{3}(l_{3}+1)}{2r^{2}\cos ^{2}\theta }
\end{equation*}
where $l_{i}$ $(i=1,2,3)$ are integer and are allowed to take the null
value, thus restoring the pure Coulomb potential.

Finally, we note that the operators 
\begin{eqnarray}
I_{1} &=&\mathbf{L}^{2}+\frac{s_{1}^{2}-\frac{1}{4}}{\sin ^{2}\theta \sin
^{2}\varphi }+\frac{s_{2}^{2}-\frac{1}{4}}{\sin ^{2}\theta \cos ^{2}\varphi }%
+\frac{s_{3}^{2}-\frac{1}{4}}{\cos ^{2}\theta }  \label{I_1_2} \\
I_{2} &=&L_{z}^{2}+\frac{s_{1}^{2}-\frac{1}{4}}{\sin ^{2}\varphi }+\frac{%
s_{2}^{2}-\frac{1}{4}}{\cos ^{2}\varphi }  \notag
\end{eqnarray}
where $\mathbf{L}^{2}$ and $L_{z}^{2}$ are the square of ``angular
momentum''\ and of its projection on the third axis, commute with the
Hamiltonian. These integrals of motion are related to the Casimir operators
of $SO\left( 6\right) $ and its $SO\left( 4\right) $ subgroup in the sense
that 
\begin{equation}
I_{1}=\left. C^{SO\left( 6\right) }\right| _{\mathcal{H}_{s_{1}s_{2}s_{3}}}%
\;,\;I_{2}=\left. C^{SO\left( 4\right) }\right| _{\mathcal{H}%
_{s_{1}s_{2}s_{3}}}\;.  \label{I_1_2_restr}
\end{equation}
where 
\begin{eqnarray*}
C^{SO\left( 6\right) } &=&-\left( \frac{1}{\sin \theta }\frac{\partial }{%
\partial \theta }\sin \theta \frac{\partial }{\partial \theta }+\frac{1}{%
\sin ^{2}\theta }\frac{\partial ^{2}}{\partial \varphi ^{2}}\right) \\
&&-\frac{1}{\sin ^{2}\theta \sin ^{2}\varphi }\left( \frac{1}{4}+\frac{%
\partial ^{2}}{\partial \alpha _{1}^{2}}\right) -\frac{1}{\sin ^{2}\theta
\cos ^{2}\varphi }\left( \frac{1}{4}+\frac{\partial ^{2}}{\partial \alpha
_{2}^{2}}\right) -\frac{1}{\cos ^{2}\theta }\left( \frac{1}{4}+\frac{%
\partial ^{2}}{\partial \alpha _{3}^{2}}\right)
\end{eqnarray*}
and 
\begin{equation*}
C^{SO\left( 4\right) }=-\frac{\partial ^{2}}{\partial \varphi ^{2}}-\frac{1}{%
\sin ^{2}\varphi }\left( \frac{1}{4}+\frac{\partial ^{2}}{\partial \alpha
_{1}^{2}}\right) -\frac{1}{\cos ^{2}\varphi }\left( \frac{1}{4}+\frac{%
\partial ^{2}}{\partial \alpha _{2}^{2}}\right)
\end{equation*}

\subsection{Bound states}

The bound state spectrum can now be easily obtained if we note that the
eigenvalue of the Casimir operator $C$ of the potential groups $SO\left(
7\right) $ is $j\left( j+5\right) $. We then find 
\begin{equation}
E=-\frac{\gamma ^{2}}{2\left( j+\frac{5}{2}\right) ^{2}}\;  \label{E_b}
\end{equation}
where $j=s_{1}+$ $s_{2}+s_{3}+2k_{1}+2k_{2}+n,\quad \left(
k_{1},k_{2},n=0,1,2,\ldots \right) $. It is easy to check that states (\ref
{E_b}) have degeneracy $\frac{d\left( d+1\right) }{2}$, where $d=\left[ 
\frac{j-s_{1}-s_{2}-s_{3}}{2}\right] +1$, and $\left[ q\right] $ is the
largest integer less than or equal to $q$.

We give for reference the expression of the bound-state wave functions 
\begin{equation}
\psi \left( x\right) =\mathcal{R}_{jl}\left( r\right) \mathcal{Y}_{lM}\left(
\theta ,\varphi \right) ,  \label{psi_b}
\end{equation}
where $\mathcal{R}_{jl}\left( r\right) $ is the radial part of the wave
function, while $\mathcal{Y}_{lM}\left( \theta ,\varphi \right) $ is the
angular part of it : \ 
\begin{equation}
\mathcal{R}_{jl}\left( r\right) =cu^{l+\frac{3}{2}}e^{-\frac{u}{2}%
}L_{n}^{2l+4}\left( u\right) \;,\;u=2\gamma r/\left( j+\frac{5}{2}\right)
\label{R_jl}
\end{equation}
with $n=j-l$ $\left( n=0,1,2,\ldots \right) $, 
\begin{equation}
c=\left( \frac{2\gamma }{j+\frac{5}{2}}\right) ^{3}\left[ \frac{\Gamma
\left( j-l+1\right) }{2\left( j+\frac{5}{2}\right) \Gamma \left(
j+l+5\right) }\right] ^{\frac{1}{2}}  \label{c}
\end{equation}
and 
\begin{equation}
\begin{array}{c}
\mathcal{Y}_{lM}\left( \theta ,\varphi \right) =\chi \sin ^{m+1}\theta \cos
^{s_{3}+\frac{1}{2}}\theta \sin ^{s_{1}+\frac{1}{2}}\varphi \cos ^{s_{2}+%
\frac{1}{2}}\varphi \\ 
\times P_{k_{1}}^{\left( m+1,s_{3}\right) }\left( \cos 2\theta \right)
P_{k_{2}}^{\left( s_{1},s_{2}\right) }\left( \cos 2\varphi \right)
\end{array}
\label{Y_lm}
\end{equation}
with $2k_{1}=l-m-s_{3},2k_{2}=m-s_{1}-s_{2}\quad \left(
k_{1},k_{2}=0,1,2,,\ldots \right) $ and 
\begin{eqnarray}
\chi &=&\left[ \frac{\Gamma \left( \frac{1}{2}\left( l+m+s_{3}+4\right)
\right) \Gamma \left( \frac{1}{2}\left( l-m-s_{3}+2\right) \right) \Gamma
\left( \frac{1}{2}\left( m+s_{1}+s_{2}+2\right) \right) }{\Gamma \left( 
\frac{1}{2}\left( l+m-s_{3}+4\right) \right) \Gamma \left( \frac{1}{2}\left(
l-m+s_{3}+2\right) \right) \Gamma \left( \frac{1}{2}\left(
m+s_{1}-s_{2}+2\right) \right) }\right] ^{\frac{1}{2}}  \notag \\
&&\times \left[ \frac{\Gamma \left( \frac{1}{2}\left( m-s_{1}-s_{2}+2\right)
\right) }{\Gamma \left( \frac{1}{2}\left( m-s_{1}+s_{2}+2\right) \right) }%
\left( 2l+4\right) \left( 2m+2\right) \right] ^{\frac{1}{2}}
\end{eqnarray}

Here, \ $L_{n}^{\alpha }$ and $P_{n}^{\left( \alpha ,\beta \right) }$ are
Laguerre and Jacobi polynomials, respectively. It is worth noting that (see
Appendix ) the $\mathcal{Y}_{lM}$ functions are related to $5$-dimensional
spherical harmonics $Y_{lM}\left( n\right) $\ (see Section 10.5 of \cite
{VK93}) in polyspherical coordinates, while $\mathcal{R}_{jl}$\ is related
to the radial part of the $6$-dimensional Coulomb wave function \cite{Ni79}
as 
\begin{equation*}
\mathcal{R}_{jl}\left( r\right) =r^{\frac{3}{2}}\mathcal{R}%
_{jl}^{Coul}\left( r\right)
\end{equation*}

\subsection{Scattering states}

Once the group structure of the problem has been recognized, the associated $%
S$ matrix can be computed by using Eqs. (\ref{assoc_1}-\ref{int_2}). This
requires knowledge of matrices $\left\langle l^{\prime }M^{\prime
}\left\vert A\right\vert lM\right\rangle $\ that intertwine Weyl-equivalent
representations of $SO\left( 6,1\right) $ in the bases corresponding to the $%
SO\left( 6,1\right) \supset SO\left( 6\right) \supset SO\left( 4\right)
\times SO\left( 2\right) \supset SO\left( 2\right) \times SO\left( 2\right)
\times SO\left( 2\right) $ reduction. One has (see Appendix) 
\begin{equation}
\left\langle l^{\prime }M^{\prime }\left\vert A\right\vert lM\right\rangle
=A_{l}\delta _{ll^{\prime }}\delta _{MM^{\prime }}\;,  \label{A_mat_el}
\end{equation}
where 
\begin{equation}
A_{l}=\frac{\Gamma \left( \frac{5}{2}+{\normalsize i}\rho +l\right) }{\Gamma
\left( \frac{5}{2}-{\normalsize i}\rho +l\right) }\;.  \label{A_l}
\end{equation}

According to this, we have 
\begin{equation}
S\left( \theta ,\varphi ;\theta ^{\prime },\varphi ^{\prime }\right)
=\sum\limits_{lM}A_{l}\mathcal{Y}_{lM}\left( \theta ,\varphi \right) 
\mathcal{Y}_{lM}^{\ast }\left( \theta ^{\prime },\varphi ^{\prime }\right)
\;.  \label{S_thetaphi}
\end{equation}

Thus, the scattering amplitude, $f\left( \theta ,\varphi ;\theta ^{\prime
},\varphi ^{\prime }\right) $, is defined by 
\begin{equation}
f\left( \theta ,\varphi ;\theta ^{\prime },\varphi ^{\prime }\right) =\frac{%
2\pi }{{\normalsize i}p}\sum\limits_{lM}\left( A_{l}-1\right) \mathcal{Y}%
_{lM}\left( \theta ,\varphi \right) \mathcal{Y}_{lM}^{\ast }\left( \theta
^{\prime },\varphi ^{\prime }\right) \;.  \label{sc_ampl}
\end{equation}

Since 
\begin{equation*}
\sum\limits_{lM}\mathcal{Y}_{lM}\left( \theta ,\varphi \right) \mathcal{Y}%
_{lM}^{\ast }\left( \theta ^{\prime },\varphi ^{\prime }\right) =\delta
\left( \cos \theta -\cos \theta ^{\prime }\right) \delta \left( \varphi
-\varphi ^{\prime }\right)
\end{equation*}
we can omit unity in the brackets of formula (\ref{sc_ampl}) when $\theta
\neq \theta ^{\prime }$, $\varphi \neq \varphi ^{\prime }$, leaving 
\begin{equation}
f\left( \theta ,\varphi ;\theta ^{\prime },\varphi ^{\prime }\right) =\frac{%
2\pi }{{\normalsize i}p}\sum\limits_{lM}A_{l}\mathcal{Y}_{lM}\left( \theta
,\varphi \right) \mathcal{Y}_{lM}^{\ast }\left( \theta ^{\prime },\varphi
^{\prime }\right) \;.  \label{sc_ampl_1}
\end{equation}

Moreover, formulas (\ref{exp}),(\ref{add}) and (\ref{harm}) imply the
following integral representation of the scattering amplitude 
\begin{eqnarray}
f\left( \theta ,\varphi ;\theta ^{\prime },\varphi ^{\prime }\right) &=&%
\frac{2\pi }{{\normalsize i}p}\eta \sqrt{b}\int\limits_{0}^{2\pi
}\int\limits_{0}^{2\pi }\int\limits_{0}^{2\pi }\left( 1-a\sin \theta \sin
\theta ^{\prime }-\cos \theta \cos \theta ^{\prime }\cos \alpha _{3}\right)
^{-\frac{5}{2}-{\normalsize i}\rho }  \notag  \label{ampl_i} \\
&&\times \exp \left( -is_{1}\alpha _{1}-is_{2}\alpha _{2}-is_{3}\alpha
_{3}\right) d\alpha _{1}d\alpha _{2}d\alpha _{3}
\end{eqnarray}
where 
\begin{equation}
a=\sin \varphi \sin \varphi ^{\prime }\cos \alpha _{1}+\cos \varphi \cos
\varphi ^{\prime }\cos \alpha _{2}  \label{a}
\end{equation}
and 
\begin{equation}
b=\sin \theta \sin \theta ^{\prime }\sin \varphi \sin \varphi ^{\prime }
\label{b}
\end{equation}

\section{Conclusions and outlook}

We have shown in the present work, based on the potential group approach,
how a non-central extension of the Coulomb Hamiltonian, considered in the
literature as an example of maximal superintegrability, can be worked out in
a fully analytic way, with bound states described by most degenerate
representations of $SO\left( 7\right) $\ and scattering states by most
degenerate representations of $SO(6,1)$. The subfamily of the generalized
Coulomb problem described in the present work does not include the pure
Coulomb potential: in order to restore it, the symmetries could be enlarged
to $SO(10)$\ and $SO(9,1)$, respectively.

The generation of solvable non-central potentials via the potential group
approach is quite general and not limited to the orthogonal and
pseudo-orthogonal groups of interest to the Coulomb problem. An example of a
non-central extension of the harmonic oscillator with $U\left( 4\right) $\
symmetry has been discussed in Ref.\cite{KE06}, while a non-central
extension of the null potential with $E\left( 4\right) $\ symmetry has been
worked out in Ref.\cite{KE07}. Other cases of physical interest with more
complicated symmetries will be considered for future work.

\section{Appendix: Calculation of the matrix elements of $A$}

Here we calculate the matrix elements of $A$ which intertwine
Weyl-equivalent representations of $SO(6,1)$\ or $\mathfrak{so}(6,1)$ in the
bases corresponding to $SO(6,1)\supset SO(6)\supset SO\left( 4\right) \times
SO\left( 2\right) \supset SO\left( 2\right) \times SO\left( 2\right) \times
SO\left( 2\right) $ reduction. We find it expedient to use, for this
purpose, equation (\ref{int_2}).

We shall start with the fact that the most degenerate principal series
representations of $SO(6,1)$ can be realized on $\mathcal{L}_{2}\left(
S^{5}\right) $ (see Section 9.2.1 of \cite{VK93}) 
\begin{equation}
U_{j}\left( g\right) f\left( n\right) =\left( \omega _{g}\right) ^{j}f\left(
n_{g}\right) \;,\quad n\in S^{5}  \label{U_g1}
\end{equation}
where 
\begin{equation*}
\omega _{g}=\sum_{i=1}^{6}g_{7i}^{-1}n_{i}+g_{77},\quad \left( n_{g}\right)
_{k}=\frac{\sum_{i=1}^{6}g_{ki}^{-1}n_{i}+g_{k7}}{%
\sum_{i=1}^{6}g_{7i}^{-1}n_{i}+g_{77}}
\end{equation*}

The operator $A$ defined by 
\begin{equation}
\left( Af\right) \left( n\right) =\int K\left( n,n^{\prime }\right) f\left(
n^{\prime }\right) dn^{\prime }  \label{Af}
\end{equation}
intertwines representations $j$ and $-5-j$, if 
\begin{equation}
K\left( n_{g},n_{g}^{\prime }\right) =\left( \omega _{g}\right) ^{5+j}\left(
\omega _{g}^{\prime }\right) ^{5+j}K(n,n^{\prime })\;.  \label{funct}
\end{equation}

The kernel, $K$ , is uniquely determined by Eq. (\ref{funct}) up to a
constant and is given by 
\begin{equation}
K(n,n^{\prime })=\eta \left( 1-n\cdot n^{\prime }\right) ^{-5-j}\;.
\label{kern}
\end{equation}
with 
\begin{equation}
\eta =2^{-\frac{5}{2}+{\normalsize i}\rho }\frac{\Gamma \left( \frac{5}{2}+%
{\normalsize i}\rho \right) }{\pi ^{\frac{5}{2}}\Gamma \left( -{\normalsize i%
}\rho \right) }
\end{equation}
With this factor the operator $A$\ becomes unitary for $j=-\frac{5}{2}+i\rho 
$ (see equation (\ref{int_m}) ).

Taking into account the fact that $5$-dimensional spherical harmonics $%
Y_{lM} $ of degree $l$ \cite{VK93} forms a bases in $\mathcal{L}_{2}\left(
S^{5}\right) $, corresponding to above\ reduction, we have the following
integral representation for the matrix elements of $A$%
\begin{equation}
\left\langle l^{\prime }M^{\prime }\right\vert A\left\vert lM\right\rangle
=\int \mathcal{K}\left( n,n^{\prime }\right) Y_{l^{\prime }M^{\prime
}}^{\ast }\left( n^{\prime }\right) Y_{lM}\left( n\right) dndn^{\prime }\;.
\label{S_lK}
\end{equation}
where $dn=\sin ^{3}\theta \cos \theta \sin \varphi \cos \varphi d\theta
d\varphi d\alpha _{3}d\alpha _{2}d\alpha _{1}$ for $n$\ as in (\ref
{polar_coord}) and 
\begin{equation}
Y_{lM}\left( n\right) =\mathcal{Y}_{lM}\left( \theta ,\varphi \right)
\prod_{j=1}^{3}\frac{1}{\sqrt{2\pi }}e^{{\normalsize i}s_{j}\alpha _{j}}.
\label{harm}
\end{equation}

By using the expansion 
\begin{equation}
\eta \left( 1-n\cdot n^{\prime }\right) ^{-\frac{5}{2}-{\normalsize i}\rho }=%
\frac{1}{2\pi ^{3}}\sum\limits_{v=0}^{\infty }\left( \nu +2\right) \frac{%
\Gamma \left( \frac{5}{2}+{\normalsize i}\rho +\nu \right) }{\Gamma \left( 
\frac{5}{2}-{\normalsize i}\rho +\nu \right) }C_{v}^{2}\left( n\cdot
n^{\prime }\right) \;,  \label{exp}
\end{equation}
we have 
\begin{equation}
\left\langle l^{\prime }M^{\prime }\right\vert A\left\vert lM\right\rangle
=A_{l}\delta _{ll^{\prime }}\delta _{MM^{\prime }}  \label{int_m}
\end{equation}
with 
\begin{equation}
A_{l}=\frac{\Gamma \left( \frac{5}{2}+{\normalsize i}\rho +l\right) }{\Gamma
\left( \frac{5}{2}-{\normalsize i}\rho +l\right) }
\end{equation}
In arriving at equation (\ref{int_m}) we have used the addition formula 
\begin{equation}
C_{\nu }^{2}\left( n\cdot n^{\prime }\right) =\frac{2\pi ^{3}}{\nu +2}%
\sum\limits_{M}Y_{\nu M}\left( n\right) Y_{\nu M}^{\ast }\left( n^{\prime
}\right)  \label{add}
\end{equation}

\end{document}